\documentclass{article}
\usepackage{waspaa21,amsmath,graphicx,url,times, subcaption}
\usepackage{color}
\usepackage{tabularx}
\usepackage{arydshln}
\usepackage{subcaption}
\usepackage{amssymb}
\usepackage{textcomp}
\usepackage{setspace}
\usepackage{enumitem}
\usepackage{siunitx}
\usepackage{hyperref}
\usepackage{caption}
\usepackage{subcaption}

\DeclareMathAlphabet{\pazocal}{OMS}{zplm}{m}{n}
\newcommand{\Loss}{\pazocal{L}}

\input{def.set}

\title{HARP-Net: Hyper-Autoencoded Reconstruction Propagation\\for Scalable Neural Audio Coding}

\name{Darius Petermann,$^{1}$
      Seungkwon Beack,$^{2}$
      Minje Kim,$^{1}$\sthanks{This work was supported by Institute for Information \& communications Technology Promotion (IITP) grant funded by the Korea government (MSIT) (2017-0-00072, Development of Audio/Video Coding and Light Field Media Fundamental Technologies for Ultra Realistic Tera-Media).}}
\address{$^1$ Indiana University, Department of Intelligent Systems Engineering, Bloomington, IN  47408, USA\\              
         $^2$ Electronics and Telecommunications Research Institute, Daejeon 34129, South Korea\\
        \texttt{daripete@iu.edu, skbeack@etri.re.kr, minje@indiana.edu}
}

\begin{document}

\ninept
\maketitle

\begin{sloppy}

\begin{abstract}
% Audio coding has been an active research area for the past few decades, while data-driven machine learning approaches have not been widely studied. 
% Based on the recent advancement in neural speech coding, we identify the needs to 
We propose a novel autoencoder architecture that improves the architectural scalability of general-purpose neural audio coding models. 
% To this end, we propose to improve the common autoencoding architecture, whose bottleneck layer divides the model into the encoder and decoder parts. 
An autoencoder-based codec employs quantization to turn its bottleneck layer activation into bitstrings, a process that hinders information flow between the encoder and decoder parts. To circumvent this issue, we employ additional skip connections between the corresponding pair of encoder-decoder layers. The assumption is that, in a mirrored autoencoder topology, a decoder layer reconstructs the intermediate feature representation of its corresponding encoder layer. Hence, any additional information directly propagated from the corresponding encoder layer helps the reconstruction. We implement this kind of skip connections in the form of additional autoencoders, each of which is a small codec that compresses the massive data transfer between the paired encoder-decoder layers. We empirically verify that the proposed hyper-autoencoded architecture improves perceptual audio quality compared to an ordinary autoencoder baseline. 

% is the only part quantized into bitstream,     Audio coding, a task consisting in reducing the amount of data needed in audio signals representation, has seen substantial improvement with the recent advent of deep learning, opening doors to neural network-based coding approaches. In this paper we propose a novel neural audio coding system revolving around the U-Net architecture. Adding up to previous work integrating the quantization part as a single step as part of the architecture bottleneck, our model integrates intermediate, layer-wise auto-encoded skip connections, each of which passes quantized feature maps from the encoding to the decoding paths. We call this propagation method "Hyper-Autoencoded Reconstruction Propagation", or "HARP-NET". We demonstrate the efficacy of our approach and improvements added by the intermediate AEs towards sound quality through objective and subjective evaluations.
\end{abstract}


\section{Introduction}

Data compression is an essential aspect of information and communication systems nowadays. Its main aim is to reduce the bitrate by eliminating the redundancy of the data present in the signal by mapping the raw data samples in the original high precision representation (e.g., single-precision floating points) into a compact discrete representation (e.g., a bitstring). In this work, we focus on the audio coding applications where the reconstructed signal on the receiver side is allowed to have reconstruction error once its perceptual quality is above the desired level. While in speech communication low bitrate codecs can achieve the common \textit{intelligibility} goal, as for music signals, the required bitrate tends to be much higher since the user's listening experience can be deteriorated even by a subtle perceptual degradation. Hence, scalability to various use cases and bitrates is an important goal in modern speech or audio codecs, such as in unified speech and audio coding (USAC) \cite{usac1, usac2}.

% The task of an audio compression systems, or codec, is two-fold: first the signal is reduced into its bitstring-code representation, a task fulfilled as part of the encoding process. At this stage the signal has reached its most compressed form. An estimation of the original uncompressed input can then be retrieved through the decoding process, where the code is converted back into its floating point representation. Originally these coding systems were manually hand-crafted towards their end-goal. Arguably the most popular and still widely used form of audio coding, MP3, is specifically suitable for musical signals as its form of compression allows for reasonable fidelity.

With the recent breakthrough in deep learning, neural speech coding emerged as a new research area. Despite their higher computationally complexity than the conventional speech codecs, such as AMR-WB \cite{BessetteB2002amrwb} and Opus \cite{ValinJM2012opus}, neural speech codecs show merits in terms of coding gain. For example, fully-convolutional autoencoders have been successfully transformed into a codec, whose bottleneck layer is quantized to produce bitstings out of waveforms \cite{KankanahalliS2018icassp}. These relatively compact waveform codecs start to compete with AMR-WB and Opus after being coupled with linear predictive coding (LPC) \cite{ZhenK2020cq}. Meanwhile, generative models, such as WaveNet \cite{OordA2016wavenet}, have proven to be effective towards speech coding reducing bitrates down to 2.4kbps, while retaining reasonable speech quality \cite{KleijnW2018wavenet, GarbaceaC2019vqvae}. In return, the system is burdened by the complex WaveNet-based decoders. A more recent approach proposed to use LPCNet \cite{ValinJ2019lpcnet}, achieving as low as 1.6kbps \cite{ValinJ2019lpcnetcoding}. Likewise, neural speech coding has advanced by encompassing traditional technology.

However, most promising advancements have been made for speech coding rather than general-purpose audio coding. For example, the WaveNet decoder is designed for very low bitrate cases, while it is too heavy for real-time decoding, e.g., for on-device music players. Meanwhile, the autoencoder-based waveform codecs display significant drawback as they rely on objective loss functions, e.g., the mean-squared error, which often result in perceptual discrepancy among decoded signals. Hence, perceptually more meaningful loss functions show promising results for neural audio coding by calibrating the loss with a psychoacoustic weighting scheme \cite{ZhenK2020spl}. In \cite{ZhenK2020spl}, the idea of cascading multiple autoencoders \cite{ZhenK2019interspeech} also show performance improvement, while the concatenation linearly increases model complexity.

\begin{figure*}[t]
     \centering
     \begin{subfigure}[b]{0.25\textwidth}
         \centering
         \includegraphics[width=\textwidth]{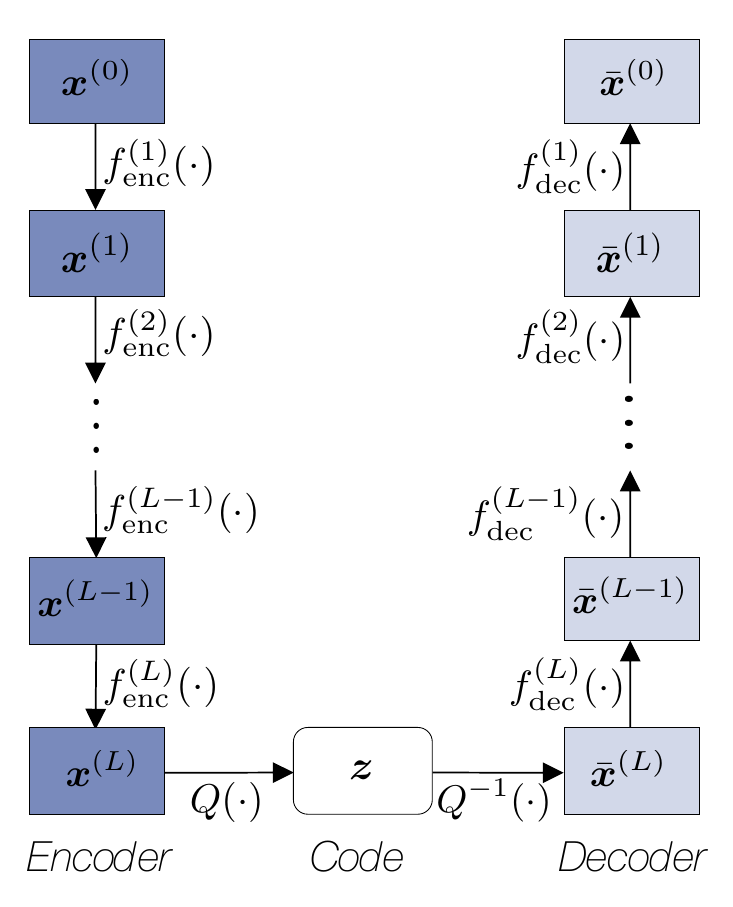}
         \caption{The baseline autoencoder}
         \label{fig:basic_ae}
     \end{subfigure}
     \hfill
     \begin{subfigure}[b]{0.25\textwidth}
         \centering
         \includegraphics[width=\textwidth]{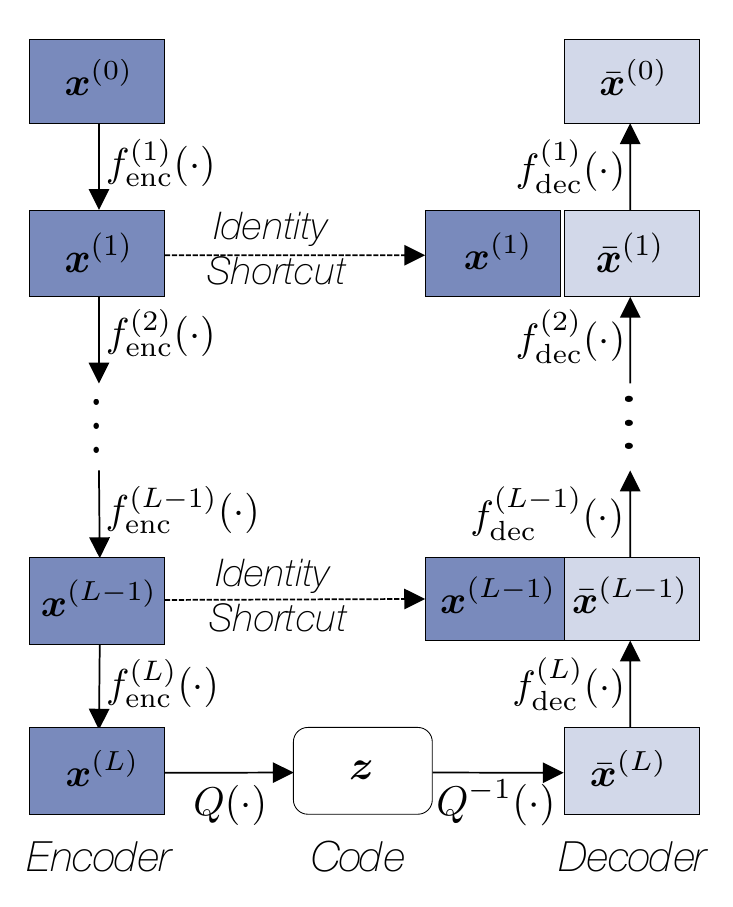}
         \caption{A hypothetical U-Net codec}
         \label{fig:unet}
     \end{subfigure}
     \hfill
     \begin{subfigure}[b]{0.46\textwidth}
         \centering
         \includegraphics[width=\textwidth]{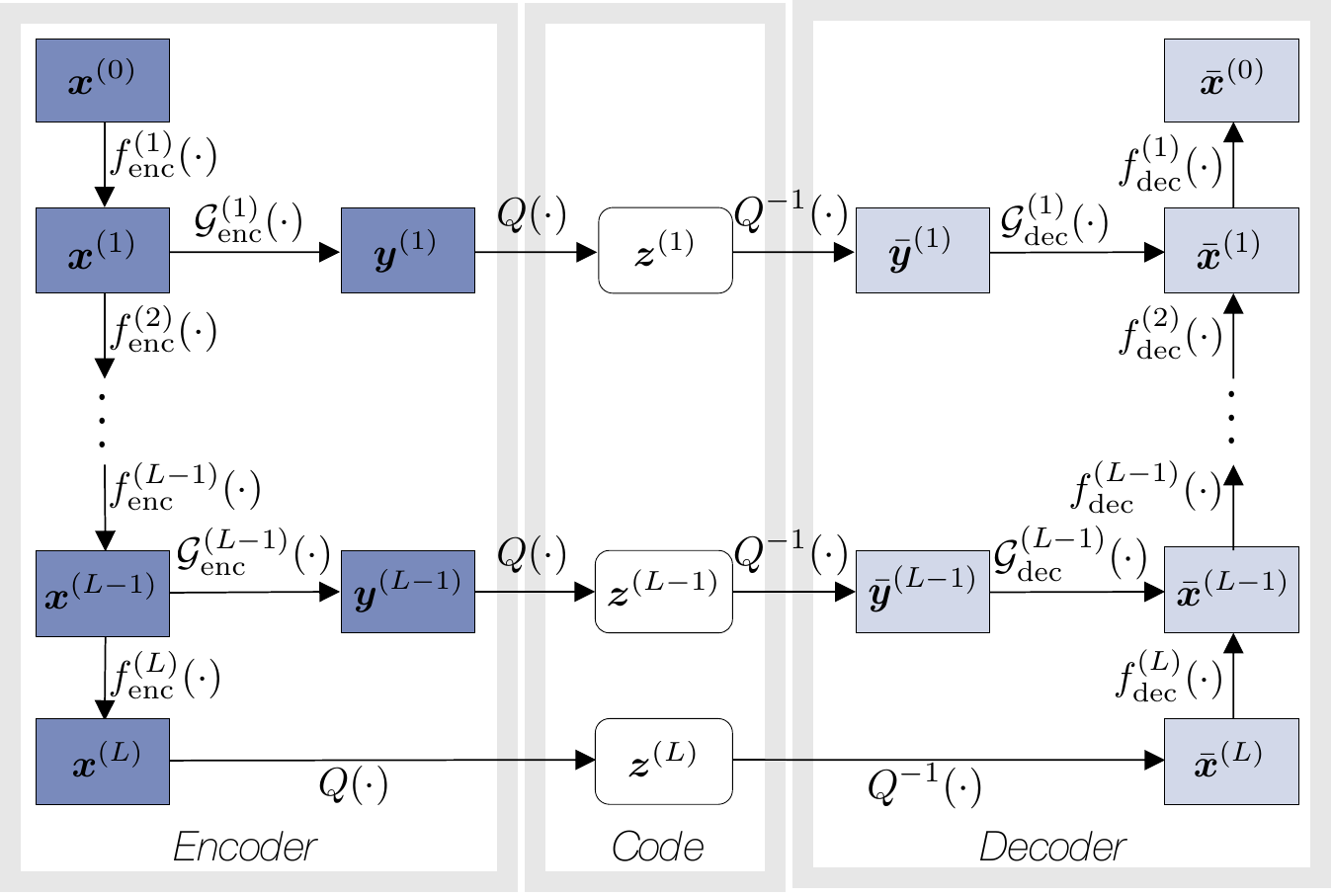}
         \caption{The proposed HARP-Net architecture}
         \label{fig:harpnet}
     \end{subfigure}
        \caption{The neural audio coding systems discussed in this work. Note that in the hypothetical U-Net codec each identity shortcut requires a substantial bitrate as each $\bx^{(l)}$ is part of code. HARP-Net compresses them further and reduces the total bitrate.}
        \label{fig:HARP-Net}
\end{figure*}

Our proposed method tackles the neural audio coding problem via architectural improvement. Our codec covers various bitrates, while its complexity is suppressed to be low. To this end, we posit a mirrored autoencoder (AE) architecture, where the feature maps on the decoder side are an approximation of the encoder's. Hence, we conjecture that direct communication between each pair of encoder and decoder layers can improve the layer-wise approximation quality of the decoder. Then, the better approximation performance is propagated to the decoder's output layer, leading to a better total reconstruction. To this end, we employ additional autoencoding paths that skip-connect the corresponding feature map pairs on the opposite sides. Since each skip AE is a small codec producing a code from its own bottleneck layer, the final bitstream is the sum of all codes from those bottleneck layers. We call the proposed architecture \textit{hyper-autoencoded reconstruction propagation network} (HARP-Net). 
HARP-Net shares the similar advantage introduced by the U-Net's skip connections \cite{RonnebergerO2015unet}: a superior reconstruction quality for autoencoding-like tasks, such as music source separation in Wave-U-Net \cite{StollerD2018waveunet}. The difference is that U-Net's identity shortcuts disqualify the architecture as a codec due to the high datarate. HARP-Net replaces the bitrate-consuming identity shortcuts with compact AEs, whose goal is to deliver the encoder feature map in a compressed form to the decoder side.

\section{Neural Audio Coding}
\subsection{Basic Autoencoders}
Figure \ref{fig:basic_ae} depicts an AE-based neural audio coding system. 
Its encoder function $\calF_\text{enc}(\cdot)$ converts the input signal into a feature vector, which is subsequently quantized into a bitstring, or the code $\bz$. Compression is achieved if $\bz$ is with a smaller number of bits than the amount necessary to represent the raw input. The decoder $\calF_\text{dec}(\cdot)$ follows to recover the original signal out of the code vector $\bz$. Hence, the AE's goal is to best predict the reconstruction $\bar\bx$ that approximates the input $\bx$ using a short enough bitstring $\bz$. We write this procedure by employing a quantizer $Q$ and its inversion:
\begin{equation}
    \bx\approx\bar\bx\leftarrow \calF_\text{dec}\circ Q^{-1}(\bz), \quad \bz\leftarrow Q\circ \calF_\text{enc}(\bx).
\end{equation}

Neural coding systems employ a deep neural network (DNN) to implement the autoencoding function. Hence, the encoder function $\calF_\text{enc}(\cdot)$ can be defined as a series of non-linear transforms, whose $l$-th layer converts its input $\bx^{(l-1)}$ into a new feature space $\bx^{(l)}$:
\begin{align}
    \bz\leftarrow\calF_\text{enc}(\bx)=Q\circ f_\text{enc}^{(L)}\circ\cdots\circ f_\text{enc}^{(2)} \circ f_\text{enc}^{(1)}(\bx),
\end{align}
where $L$ denotes the number of the encoder layers. Hence, we can also denote the raw input as  $\bx=\bx^{(0)}$. 
The encoder output $\bx^{(L)}$ goes through a quantization module $Q$ that discretizes the floating-point feature vector $\bx^{(L)}$ into a bitstring $\bz$. Furthermore, $\bz$ is losslessly compressed by using entropy coding, e.g., Huffman coding.

% , but we will not discuss about it, as it is an obvious lossless compression process. 

The de-quantization process $Q^{-1}(\bz)$ reverts the code vector $\bz$ back to the floating-point feature $\bar\bx^{(L)}$ with a certain amount of quantization error, i.e., $\bar\bx^{(L)}$ approximates $\bx^{(L)}$. Then, a series of decoder layers recover the raw signal:
\begin{equation}
    \bar\bx^{(0)}\leftarrow\calF_\text{dec}(\bx)=f_\text{dec}^{(1)}\circ\cdots\circ f_\text{dec}^{(L-1)} \circ f_\text{dec}^{(L)}\circ Q^{-1}(\bz).
\end{equation}
Note that layer indices decrease from the bottleneck to the output layer.

Although it is not common in the recent deep learning literature, a hard association between the encoder and decoder layers was popular in some earlier models. For example, in the restricted Boltzmann machines (RBM)  \cite{HintonG2006rbm}, we write the $l$-th layer transform $\bx^{(l)}\!\leftarrow\! \bW_\text{enc}^{(l)} \bx^{(l-1)} + \bb_\text{enc}^{(l)}$. Then, its corresponding decoder layer is defined with the encoder weight matrix $\bW_\text{enc}^{(l)}$, but by transposing it: $\bar\bx^{(l-1)}\!\leftarrow\!\left(\bW_\text{enc}^{(l)}\right)^\top \bar\bx^{(l)} + \bb_\text{dec}^{(l)}$. Since the transpose operation is equivalent to matrix inversion for orthonormal matrices, this association loosely implies an inversion relationship. 

The basic AE in Figure \ref{fig:basic_ae} is with a mirrored architecture, while the coupling of the encoder and decoder layers is not assumed. 
% In the proposed models though, we postulate that the weight matrix of $f_\text{enc}^{(l)}(\cdot)$ is associated with $f_\text{dec}^{(l)}(\cdot)$ for all $l$. The association means that the decoder layer reverses the transformation operation that its corresponding encoder layer conducts. 
In the proposed HARP-Net architecture, we postulate the ties between the encoder-decoder layers must be helpful for coding, although the coupling is indirect using skip connections rather than inversion. 

\subsection{Quantization and Bitrate Control}\label{sec:quant}

% <\cite{KankanahalliS2018icassp}>
% <\cite{zhen2019efficient}>
% <\cite{AgustssonE2017softmax}>

% \subsubsection{Soft-to-Hard Quantization}

The quantization function $Q$ maps a floating-point value $x^{(L)}_i$ to a pre-defined, finite set of quantization bins. Instead of post-training quantization, we include $Q$ and $Q^{-1}$ as part of the AE, so training of the AE can also suppress the quantization artifact. Since $Q$ is not a differentiable process, we employ soft-to-hard quantization \cite{AgustssonE2017softmax}, which softens the bin assignment during training. First, it estimates the probability of assigning $i$-th code to $j$-th bin, $\bP=\text{softmax}(\alpha \bS)$, where $S_{ij}$ defines the similarity (e.g. negative absolute difference) between $i$-th code value and $j$-th representative $\mu_j$. $\alpha$ controls the ``hardness" of the logits, i.e., the smaller the softer. The softmax function turns the $i$-th similarity values $\bS_{i:}$ to all $J$ quantization bins into a probability vector $\bP_{i:}$. Then, the convex combination of the soft assignment probability and bin representatives recovers the feature vector: $\tilde{x}^{(L)}_i = \sum_{j=1}^J P_{ij}\mu_j$. The decoder takes $\tilde{\bx}^{(L)}$ as input. This whole process is differentiable. 

Conversely, the test-time inference uses the hard quantization. First, from $\bS$, the closest bin index for all scalar codes forms the discrete code $\bz$, i.e., $j^*=\argmax_j \bS_{ij}$. Note that we reduce the gap between the soft and hard assignments by annealing $\alpha$, i.e., by gradually increasing it during training: $\lim_{\alpha\rightarrow\infty}\tilde{\bx}^{(L)}=\bar{\bx}^{(L)}$.  

% while it is still functionally similar to the test-time quantization process. making BP  Consequently, instead of assigning our floating point vector through a hard nearest neighbor method, we relax this process by using soft assignment:  This means that the system entropy approximation can be defined as the texti{partial} code assignment histogram, as follows: $p = \sum_{i=0}^{N} \sum_{J=0}^{N} S_{i,j}  \in \mathbb{R}^L$ where $N$ and $M$ each denote the feature dimensions, $L$ the code length and $S$ the soft-assignment code vectors, defined as: $S  = \text{softmax}(-\sigma \ \overline{z})$. The \textit{softmax} operator assigns each entry from its input to probabilistic vector output of positive mappings. With this soft relaxation, the entropy loss is now differentiable.

% from this histogram we can then derive an approximation of the overall system entropy, as follows: $-\sum_{l=1}^{L} p_l \ \text{log} \ p_l $. Since the $\sigma$ controls the hardness of the soft assignment $S$, the entropy approximation gets closer to the true system entropy as $\sigma \rightarrow \infty$.

% \subsubsection{Entropy}

The code $\bz$ is further compressed via Huffman coding losslessly, the entropy of $p(x^{(L)}_i)$ serves as the lower bound of the bitrate, which is unknown. As an alternative, we predict the overall frequency of $j$-th bin being selected, $p_j$, by observing the soft assignment probability $\bP$ for all training examples. Hence, the approximate entropy is $\calH=-\sum_j p_j \log p_j$, which we regularize during training to control the bitrate. For example, the encoder adjusts the distribution of $\bx^{(L)}$, so its entropy matches the target entropy, i.e., the lower bound of the per-frame bitrate.

\subsection{The Hypothetical Codec with the U-Net Architecture}

% The U-Net architecture was initially proposed in the context of medical image segmentation \cite{RonnebergerO2015unet}. The architecture was adopted towards audio applications both in the spectral \todo{cite} and time domains todo \cite{StollerD2018waveunet}. 
In the original U-Net \cite{RonnebergerO2015unet} and its variants for audio processing \cite{StollerD2018waveunet, JanssonA2017cnn_unet} skip connections pair up the encoder and decoder layers through \textit{identity shortcuts} to improve the AE's reconstruction performance. Hence, the encoder feature maps are delivered to the decoder layers intact. Figure \ref{fig:unet} depicts a hypothetical U-Net architecture for neural coding, whose decoder layer now takes as input both the corresponding encoder and decoder feature maps by concatenating them together along the channel dimension: 
\begin{equation}
    \bar\bx^{(l-1)}\leftarrow f_\text{dec}^{(l)}\left(\big[\bx^{(l)}, \bar\bx^{(l)}\big]\right).
\end{equation}

% \begin{figure}
%     \centering
%         \includegraphics[scale=0.1]{figs/unet.jpg}
%     \caption{The U-Net architecture, with added identity short cuts at each level}
%     \label{fig:unet}
% \end{figure}

The performance gain entailed by these identity shortcuts is without question. In addition to the information decoded from the binary code vector $\bz$, the decoder layer directly utilizes the encoder feature maps, which can be regarded as \textit{less abstract} feature representations. However, it is an unrealistic coding system since each identity shortcut consumes a significant amount of bits. 
% For example, if $L$ is large, the identity shorcuts will increase the data rate, even more than just sending the raw input signal to the receiver. 

\section{The Proposed HARP-Net Architecture}

We address the hypothetical U-Net codec's high data rate issue by replacing the identity shortcuts with additional AEs, which we call \textit{skip autoencoders}. These skip connections can still be considered a way to provide less processed, i.e., \textit{rawer} information to the decoder as in U-Net. In this view, since a decoder feature map $\bar{\bx}^{(l)}$ is an approximation of its corresponding encoder representation $\bx^{(l)}$, information coming directly from the encoder will help its reconstruction job. In neural coding, these skip connections can compensate the quantization artifact introduced in the bottleneck layer, i.e.,  $\bz\leftarrow Q(\bx^{(L)})$, which propagates to all decoder layers.  

% \begin{figure}

%     \centering
%         \includegraphics[scale=0.08]{figs/HARP-Net.jpg}
%     \caption{HARP-Net.}
%     \label{fig:HARP-Net}
% \end{figure}

HARP-Net employs layer-wise skip AEs $\calG^{(l)}$ to compress the data transferred through U-Net's skip connections:
\begin{equation}
    \bx^{(l)}\approx\bar\bx^{(l)}\leftarrow \calG_\text{dec}^{(l)}\circ Q^{-1}(\bz^{(l)}), \quad \bz^{(l)}\leftarrow Q\circ \calG_\text{enc}^{(l)}(\bx^{(l)}).
\end{equation}
Now a skip connection is divided into its own encoder $\calG_\text{enc}^{(l)}(\cdot)$ and decoder $\calG_\text{dec}^{(l)}(\cdot)$ parts. Hence, instead of transmitting $\bx^{(l)}$ directly, a skip AE compresses it down to $\bz^{(l)}$, a layer-specific code vector. To this end, first, the encoder defines a feature transformation function $\by^{(l)}\leftarrow\calG_\text{enc}^{(l)}(\bx^{(l)})$, which produces a a processed version of the original feature map. $\by^{(l)}$ is followed by the ordinary quantization and Huffman coding that construct the layer-wise bitstring $\bz^{(l)}$. Here, the encoder function $\calG_\text{enc}^{(l)}(\cdot)$ is a placeholder: it can employ any adequate structure, probably with multiple layers. De-quantization recovers $\bar\by^{(l)}$ from $\bz^{(l)}$. Then, the decoder recovers the feature map $\bx^{(l)}\approx\bar\bx^{(l)}\leftarrow\calG_\text{dec}^{(l)}(\bar\by^{(l)})$

Once again, the goal here is to recover the original feature $\bx^{(l)}$ with little information loss, using only few bits to represent $\bz^{(l)}$. The final code of the entire system is defined by concatenating all layer-wise codes including the ordinary bottleneck code at $L$-th layer: $\bz=[\bz^{(L)}; \bz^{(L-1)}; \cdots \bz^{(1)}]$.

\setlist{nosep}
HARP-Net's advantage can be summarized as follows:
\begin{itemize}[leftmargin=0in]\setlength{\itemindent}{.15in}
    \item \textit{Performance}: The additional skip connections boost the reconstruction quality on the decoder side than a single bottleneck code.
    \item \textit{Scalability}: Having $\bz^{(L)}$ as the most abstract form of representation, it is natural to perform scalable coding by employing $M$ additional skip AEs as neeeded, i.e., $\bz=[\bz^{(L)}; \bz^{(L-1)}; \cdots \bz^{(L-M)}]$. $M$ can control the bitrates.
    \item \textit{Flexibility}: The proposed method can be applied to any DNN with a mirrored autoencoding architecture. Furthermore, the skip AE's architecture is open to various choices. 
\end{itemize}

\begin{table}[t]
% \vspace{0.5cm}
% \setlength\tabcolsep{5.0pt}
% \def\arraystretch{1.0}
\begin{center}
\begin{tabular}{c c c c}
    Models & \# Layers & \# Filters & \# Params.\\
    \hline

    {Baseline 1}    
    & 10 & 30 & 218k \\
    {HARP-Net (1 Skip AE)}     
    & 12 & 24 & 216k \\ 
    
    \hdashline

    {Baseline 2}    
    & 9 & 36 & 275k \\
    {HARP-Net (2 Skip AEs)}     
    & 12 & 24 & 257k \\
    
    \hdashline
    
    {Baseline 3}     
    & 10 & 36 & 315k \\
    {HARP-Net (3 Skip AEs)}     
    & 12 & 24 & 298k \\ 
    
    \hdashline

    {Baseline 4}  
    & 10 & 38 & 350k \\
    {HARP-Net (4 Skip AEs)}     
    & 12 & 24 & 340k \\
    
    \hline
\end{tabular}
\end{center}
\caption{The network topology of variant HARP-Net versions with a varying number of skip AEs. Their associated plain AE baselines are presented with a matching number of parameters. The number of layers do not include the skip AEs while the number of parameters take them into account.}
\label{table:models}
\end{table}

\section{Experiments}
\subsection{The Experimental Setup}

We assess the performance of HARP-Net on musical signals. To this end, we use $6,000$ commercially released songs of 13 genres, totalling $5.5$ hours of audio data, which is divided into $5,700$ and $300$ songs for training and validation, respectively. $16$ songs are set aside as a test set. All signals are in mono with a sampling rate of $44.1$kHz. We use a frame size of 16384 with a hop size 32 samples. 

The proposed HARP-Net codec operates in two target bitrate setups, 24kbps and 48kbps, but they are defined in the LPC residual domain. To this end, the corresponding LPC coefficients need to be accounted for in the final system bitrate, which amount to 16kbps. Eventually, the resulting bitrates are 40kbps and 64kbps. We opt to remain in relatively high bitrates as we limit our experiment to 44.1kHz signals. We leave lower bitrate cases for future work.

In order to assess the impact of the proposed skip AEs, we train multiple model variants that are with varying number of skip AEs from $M=1$ to $4$. For a fair comparison, we assess each of the HARP-Net models against a vanilla AE with matching number of parameters. Table \ref{table:models} summarizes the various models trained for our experiments. Note that HARP-Net consists of 12 layers, each of which is with 24 1-d convolutional kernels of size $1\times 15 \times 24$. The exception is the final encoder layer, where only one kernel is used to collapse the 24 channels down to 1. Eventually, the dimension of the input frame and encoder output are the same, i.e., $\bx^{(L)}, \bx^{(0)}\in\Real^{16384}$, meaning the encoder does not perform dimension reduction. This choice is to avoid artifact commonly caused by the downsampling and upsampling processes in convolutional models \cite{ShiW2016superresolution}. Instead, we control the bitrate by regularizing the entropy of $\bx^{(L)}$ as discussed in Sec. \ref{sec:quant}.

% \subsubsection{Main AE}
% Our proposed model consists in a 12-layer convolutional deep U-Net architecture with a set number of 24 filters across each layer. At the bottleneck layer, the feature map dimensions are then reduced to the original network input size: 

% \begin{equation}
%     f^L_{enc}(x^{(L-1)}) = x^{(L)}
% \end{equation}

% where $x^{L-1} \in \mathbb{R}^{16384 \text{x} 24}$ and $x^{L} \in \mathbb{R}^{16384 \text{x} 1}$

% \darius{Ensuring that the dimensions of the code vector does not exceed the size of the original input will guarantee that the system will consume a reduced and reasonable amount of bits in regards to the system's target entropy.}

% In a traditional AE, the goal of the dimension reduction is to transform the signal into a feature space, in our case, where the entropy coding can be done effectively. One of the main drawback of this approach relates to the artifacts caused by the upsampling process. U-Net's skip connections are meant to partially overcome this issue by passing identity information across the two paths, however by \textit{encoding} these short cuts we somewhat lose this benefit. In this view we opt to diverge away from the downsampling/upsampling idea and stick to a set number of feature maps across each of the layers. 

HARP-Net's skip AEs all follow the similar architecture to the baseline AE, but in a smaller scale with only three hidden layers and 24 filters per layer. The bottleneck is once again reduced to the original input dimension before quatization. We learn $32$ cluster means per each bottleneck quantization task $\by^{(l)}$ as well as the main bottleneck $\bx^{(L)}$, leading to a $5$-bit quantization. To limit the dynamic range of the codes, we use the hyperbolic tangent function as the final activation of the encoders. 

% . We incorporate the quantization portion at the bottleneck of each of the network's AEs and define the number of learnable quantization clusters as \num{32} ($2^5$-bits system). We constraint the range of the quantization input in $Q(\text(tanh(z^{l})))$ to facilitate training but also avoid potential additional quantization error due to the larger range of values entailed by the model's input rescaling.

\subsection{Training Process}

% \darius{Our network is trained on the residual portion of a 16th-order linear prediction process, therefore one essential pre-processing step is to} 
We first compute the residual LPC signals from the entirety of our dataset along their respective LPC coefficients. The residual signals are then scaled up by $100$ to make up the low amplitudes before being fed to the networks. 
% Due to the low amplitudes of the LPC residual, our network input signal, we opt to scale the signal up by a factor of \num{100} prior to training. This procedure also compels use, during testing time, to scale the input signal up by a scaling factor and back down prior to LPC re-synthesis. 
Following the same method proposed in \cite{KankanahalliS2018icassp}, we first let the network train without quantization for about 8 epochs,
% , which is the time needed for the model to start converging. \darius{By proceeding this way we allow the network to smoothly pre-train in an non-quantized setting in order to improve its stability in the long run.} 
then introduce the soft-to-hard quantization module. We opt to progressively anneal $\alpha$ at a constant rate (\num{0.3} per epoch). 
% \darius{Prior work on this front proposed to integrate this step as part of the network's objective function and thus ultimately make $\sigma$ a trainable parameter.} 
% By excluding an additional term off the network objective functions we make the training process more manageable and stable. \darius{We start with an initial $\sigma$ value of 10 and train our model for 600 epochs. Consequently we end up with a $\sigma$ value of \num{190} near the end of the training.} 
We define our network loss function as a combination of the sum of squared error and entropy regularizer: 
\begin{equation}
    \Loss = \sum_t(x_t -\hat{x}_t)^2 + \lambda \sum_{m=0}^M\calH^{(L-m)},
\end{equation}
where $\lambda$ controls the amount of entropy regularization, which we dynamically change during the training process. We define the entropy regularizer as the sum of
% $\mbox{\fontsize{16}{20}\selectfont\(\varepsilon\)}(A_{\text{soft}})$ denotes the sum of soft assignments over all code vectors, that is, a 
estimated entropy values of all $M$ participating layers. Given that, a stronger regularization (with a large $\lambda$) suppresses the totaly entropy more and vice versa. The actual entropy control works by monitoring all $\calH^{(L-m)}$ values and by adjusting $\lambda$ accordingly until the sum reaches the target entropy.

% While perceptually-motivated losses \cite{ZhenK2020spl} and more complex LPC approaches \cite{zhen2019efficient} have proven to be beneficial, we leave the potential harmonization to future work.

% \ref{fig:entropy_loss} gives a general overview of the process and provides a mathematical representation. Incorporating the entropy as part of the loss is a way to ensures that the sample distributions over the kernels are not too sparse and are optimized for the targeted entropy.

% \begin{figure}
%     \centering
%         \includegraphics[scale=0.1]{figs/entropy_loss.jpg}
%     \caption{The network regularizer is defined as the sum over all weighted soft assignments $L$. The weights $\tau$ control the amount of contribution of each soft assignment toward the total loss $\Loss$.}
%     \label{fig:entropy_loss}
% \end{figure}

% The soft-assignment weights $\tau_{l}$ are dynamically set through the training to control the entropy of the AE's code vector $l$: by reducing $\tau_{l}$, the soft entropy for the AE's code vector $l$ is deemed to become larger, since reducing its weight signify a reduction of the contribution of the soft entropy to the total loss. On the other hand, increasing its weight will force the soft entropy to become smaller, as its contribution to the total loss increases.

% \darius{Finally, we opt to train our model on $44.1$kHz sampling rate. Although more challenging, it is a suitable rate for musical signals.}

\subsection{Experimental Results and Discussion}

% \subsubsection{Evaluation Metrics}

% As previously described, we evaluate our proposed systems given two bitrate case; a low bitrate case of 40-kbps and a high bitrate case of 64-kbps. We also evaluate their respective baseline model for comparison. After the training completed, 

As an objective measure, we first compute the signal-to-noise ratio (SNR) on the test set. While SNR values are not completely reliable, their correlation with subjective score can sometime be found high \cite{EmiyaV2011ieeeaslp}. Hence, we use SNR scores to get a general idea of the models' performance among the variants of HARP-Net. We eventually choose the most promising variant for the subjective tests. 
% Secondly we can compare these objective scores against the MUSHRA subjective listening tests scores we also conducted and observe how well these metrics potentially correlate with each other.

\begin{figure}[t]
     \centering
     \begin{subfigure}[t]{0.51\columnwidth}
         \centering
         \includegraphics[width=\textwidth]{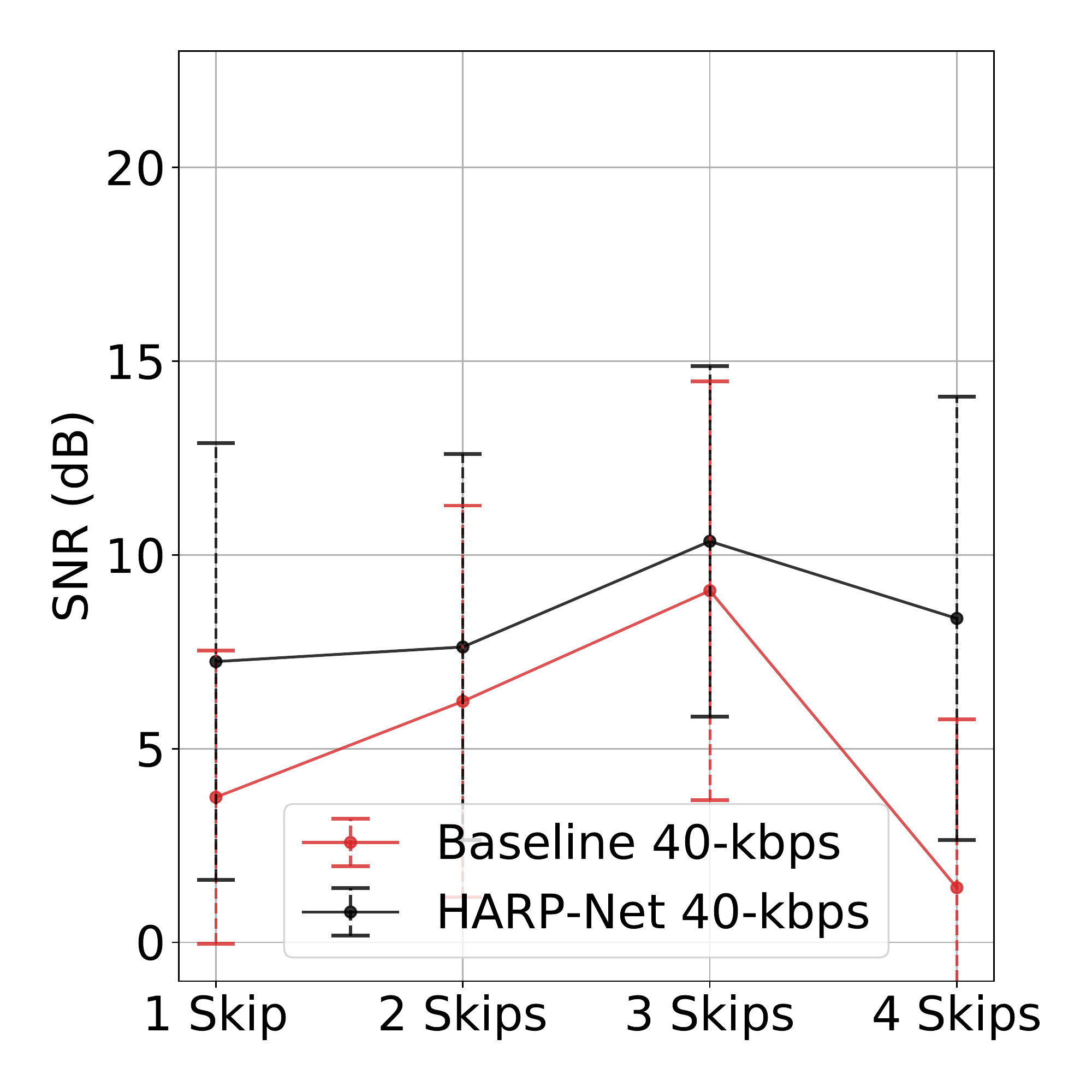}
         \caption{40kbps}
         \label{fig:snr_low}
     \end{subfigure}
    %  \hfill
     \begin{subfigure}[t]{0.481\columnwidth}
         \centering
         \includegraphics[width=\textwidth]{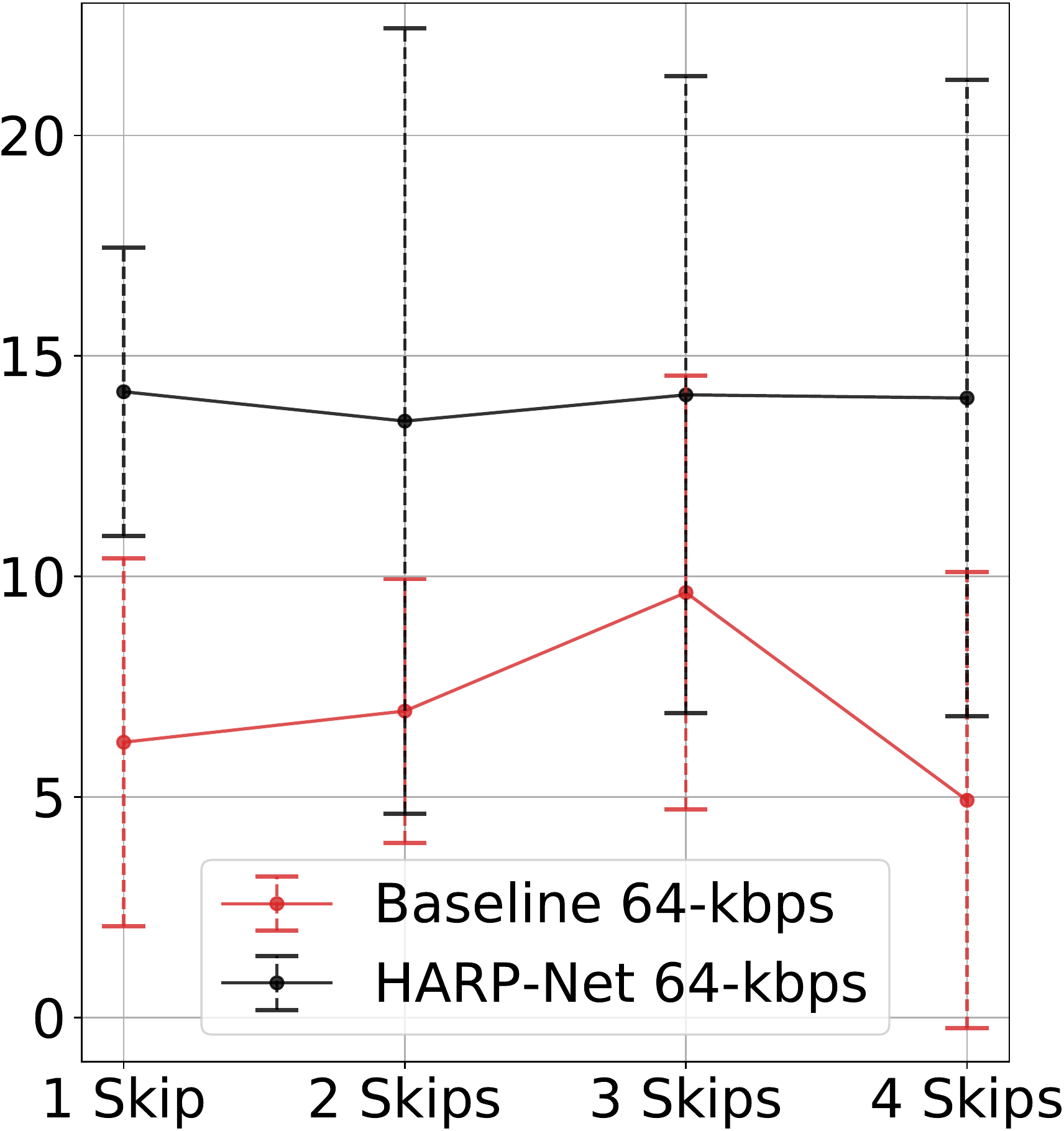}
         \caption{64kbps}
         \label{fig:snr_high}
     \end{subfigure}
    %  \hfill
    %  \begin{subfigure}[b]{0.3\textwidth}
    %      \centering
    %      \includegraphics[width=\textwidth]{graph3}
    %      \caption{$y=5/x$}
    %      \label{fig:five over x}
    %  \end{subfigure}
    %     \caption{Three simple graphs}
    %     \label{fig:three graphs}
    \caption{SNR results of HARP-Net models and their associated baselines. Note that the baselines do not have skip connections.}
    \label{fig:results_snr}
\end{figure}

\begin{figure}[t]
    \centering
        \includegraphics[scale=0.24]{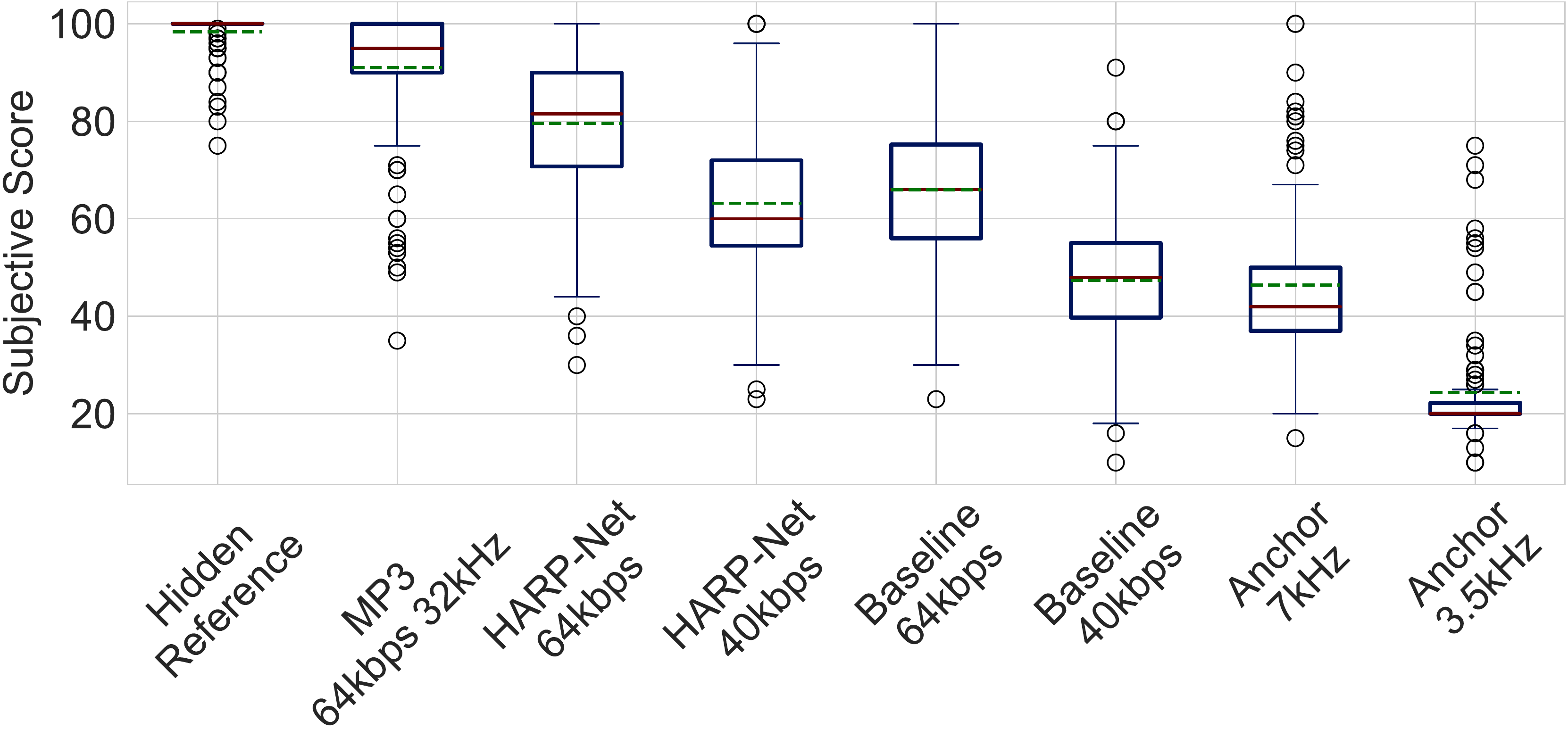}
    \caption{MUSHRA evaluation results in boxplots; green dotted line depicts the mean while the red line represent the median.}
    \label{fig:results_mushra}
\end{figure}

Fig.\ref{fig:results_snr} shows the average test SNR scores. We found that the low bitrate systems are affected more by the architectural choice, so we choose \textit{HARP-Net 3} and \textit{Baseline 3} as the best models used for subjective tests. However, note that this choice is not completely reliable due to its objective nature. Meanwhile, HARP-Net systems are not affected too much by the number of skip connections in the high bitrate cases, so we choose \textit{HARP-Net 1}. It is noticeable that the baselines' performance does not improve much by increasing target bitrate. We choose \textit{Baseline 3} as the best model for 64kbps. 

% \added{We also note the minor performance gain brought by additional skip AEs further from the bottleneck. We speculate that this could be due to the skip AE and its proximity to the bottleneck.}

The subjective test follows ITU's  MUSHRA guideline \cite{mushra}. A session consists of nine different trials, each of which includes song segments from the same test set we have set aside. We select these segments in a manner that the subjects are exposed to a wide variety of music genres, consequently covering a range of auditory qualities and aspects. Each trial includes a hidden reference along with two low-pass filtered anchors per the MUSHRA specification. The systems in comparison are two HARP-Net models in two bitrates and their corresponding baselines that are chosen based on the SNR scores. Finally, we also include the MP3 decoded signals using Adobe Audition \textregistered (licensed from Fraunhofer IIS and Thomson) at 64kbps. Note, however, that the MP3 signals are downsampled to $32$kHz by default. We recruit 12 audio experts, but excluded one of them who did not meet the post-screening guideline \cite{mushra}.

% We can observe similar trends from the MUSHRA results, with our proposed models surpassing both of their baseline, by a greater margin for the 64-kbps scenario. In this case, however, our 40-kbps model \textit{surpasses} the 64-kbps baseline model by only a small margin. 

In Fig.\ref{fig:results_mushra}, both of our proposed systems (HARP-Net 64kbps and 40kbps) outperform their baseline counterparts. Moreover, we observe that the 40kbps HARP-Net performs within the same score span as the 64kbps baseline. Lastly, we note that HARP-Net 64kbps results lay close to the MP3 performance. Given that our LPC part is not optimal and our models cover wider bandwidth, i.e., 44.1kHz instead of 32kHz, the \mbox{HARP-Net} results are promising. 

\section{Conclusions}

In this work we proposed a novel, lightweight, neural audio coding system, HARP-Net. It is defined as a mirrored AE with interconnected encoder-decoder layers via additional skip AEs. We found that those additional skip AEs create information paths that circumvent the lossy bottleneck quantization process, thus improving the total AE reconstruction performance. The skip AEs are carefully designed to work as a mini codec not to consume too much bitrates. Listening tests verified that the subjects prefer HARP-Net to the basic AE-based codecs. 
% \added{While this work aimed to confirm that our newly proposed architecture was indeed effective, there exist various ways to bring the performance of our initial system up.} 
In the future, we will improve the LPC module's bitrate, which can go down to 2.4kbps, and incorporate perceptually-motivated losses as proposed in \cite{ZhenK2020spl}. Source codes and sound examples are available at \url{https://saige.sice.indiana.edu/research-projects/harp-net}.

\bibliographystyle{IEEEtran}
\bibliography{mjkim.bib}
%
% or list them by yourself
% \begin{thebibliography}{9}
% 
% \bibitem{waspaa21web}
%   \url{http://www.waspaa.com}.
%
% \bibitem{IEEEPDFSpec}
%   {PDF} specification for {IEEE} {X}plore$^{\textregistered}$,
%   \url{http://www.ieee.org/portal/cms_docs/pubs/confstandards/pdfs/IEEE-PDF-SpecV401.pdf}.
%
% \bibitem{PDFOpenSourceTools}
%   Creating high resolution {PDF} files for book production with 
%   open source tools, 
%   \url{http://www.grassbook.org/neteler/highres_pdf.html}.
%
% \bibitem{eWilliams1999}
% E. Williams, \emph{Fourier Acoustics: Sound Radiation and Nearfield Acoustic
%   Holography}. London, UK: Academic Press, 1999.
% 
% \bibitem{ieeecopyright}
%   \url{http://www.ieee.org/web/publications/rights/copyrightmain.html}.
%
% \bibitem{cJones2003}
% C. Jones, A. Smith, and E. Roberts, ``A sample paper in conference
%   proceedings,'' in \emph{Proc. IEEE ICASSP}, vol. II, 2003, pp. 803--806.
% 
% \bibitem{aSmith2000}
% A. Smith, C. Jones, and E. Roberts, ``A sample paper in journals,'' 
%   \emph{IEEE Trans. Signal Process.}, vol. 62, pp. 291--294, Jan. 2000.
% 
% \end{thebibliography}

\end{sloppy}
\end{document}